\documentclass[aps,prper,reprint,showpacs,longbibliography]{revtex4-2}

\usepackage[T1]{fontenc}
\usepackage[utf8]{inputenc}
\usepackage{geometry}
\usepackage{graphicx}
\usepackage[export]{adjustbox}
\usepackage{times}
\usepackage{xcolor}
\usepackage{array}
\usepackage{url}
\usepackage{enumerate}
\usepackage{amsmath}
\usepackage{amssymb}
\usepackage{tikz}
\usepackage{graphicx}
\usepackage{multirow}
\usepackage{tcolorbox}
\usepackage{float}
\usepackage{booktabs}
\usepackage{ragged2e}
\usepackage[normalem]{ulem}
\usepackage{setspace}
\usepackage{caption}
\usepackage[english]{babel}
\usepackage[autostyle, english = american]{csquotes}
\MakeOuterQuote{"}
\usepackage{hyperref}
\hypersetup{
    colorlinks=true,
    citecolor=blue,
    linkcolor=blue,
    urlcolor=black}

\makeatletter

\def\section{\@startsection{section}{1}{\z@}%
  {-6ex plus -0.4ex minus -0.2ex}
  {2ex plus 0.2ex}
  {\centering\normalfont\normalsize\bfseries}
}

\def\subsection{\@startsection{subsection}{2}{\z@}%
  {-1.5ex plus -0.3ex minus -0.2ex}
  {1ex plus 0.1ex}
  {\centering\normalfont\normalsize\bfseries}
}

\def\subsubsection{\@startsection{subsubsection}{2}{\z@}%
  {-1.5ex plus -0.3ex minus -0.2ex}
  {1ex plus 0.1ex}
  {\centering\normalfont\normalsize\itshape}
}
\makeatother

\begin{document}

\onecolumngrid

\begin{center}

{\Large\bfseries Exploring Students' Perceptions of Using Generative AI-Assisted Problem Posing\par}

\vspace{0.75cm}

{\large Lindsay Dawson\par}
\textit{Department of Physics and Astronomy\\
Purdue University\\
West Lafayette, IN 47907, USA}

\vspace{0.75cm}

{\large N. Sanjay Rebello\par}
\textit{Department of Physics and Astronomy / Department of Curriculum \& Instruction\\
Purdue University\\
West Lafayette, IN 47907, USA}

\vspace{0.5cm}

\begin{minipage}{0.9\textwidth}
Problem posing, a pedagogical practice that asks students to generate novel problems or meaningful variations to problems encountered, supports transfer of learning and strengthens problem-solving skills in physics. However, generating physics problems can be challenging, particularly for novice learners. This study investigates students' perceptions of an approach to facilitate their use of Generative AI in ways that maximize their benefits and limit risks. Students were introduced to Generative AI-assisted problem posing as a self-study technique. This study utilizes a phenomenological approach to investigate students’ perceptions on how training shaped AI interactions, attitudes towards problem posing with Generative AI, and how students view its incorporation into personal study practices. Results of this study suggest that students perceived a positive change in their interactions with Generative AI after receiving training on prompt engineering techniques. This study also reveals that students hold generally positive views towards the problem-posing technique, with a smaller subset of students showing hesitations towards using Generative AI. These results lay the foundation to introduce and employ training methods for Generative AI more widely and to continue to incorporate Generative AI into structured study techniques, like problem posing.

\end{minipage}

\vfill
\end{center}

\clearpage
\twocolumngrid

\section{Introduction}

\par Recently physics education research has begun to focus on  Generative Artificial Intelligence (Gen AI) and how it can be implemented  appropriately in educational settings \cite{ding2023virtualtutor, kilde2025labpartner,tang2025TA, zhu2024humangenerativeaicollaborativeproblem, mirza2025hype}. In recent years, students have been drawn to Gen AI tools as a source for on-demand help with difficult assignments, explaining concepts, and exploring new ideas \cite{wang2025crutch}. At the same time, Gen AI poses risks to student's autonomy in problem-solving settings, where students might be inclined to blindly trust AI generated outputs or over-rely on them, if not properly informed of tools to mitigate these risks \cite{wang2025crutch}. Research  suggests that prompt engineering, the process of tailoring your ask to Gen AI, is a key to enhancing personal engagement and effectiveness of generated outputs \cite{chen2025prompteng, wang2025crutch}. However, there is limited research on implementing and educating students about Gen AI for more structured study habits outside of the classroom. 
\par This study investigates students' perceptions of an approach to facilitate their use of Gen AI in ways that increase their benefits and limit risks. In this study, we introduce students to Generative AI-assisted problem posing as a self-study technique. Problem posing is the process of generating a meaningful variation of a solved problem to generate a new problem \cite{silver1994problemposing}. This method has been shown to improve student problem-solving skills and support transfer of learning \cite{ergun2016probposeonsolve, akben2020probposmeta, singer2013probposemath, cai2015probposemath}. In this study, we train students on the use of Gen AI and prompt engineering in problem posing. We examine students' perceptions of the impact of training on their interactions with Gen AI and the usefulness of Gen AI-assisted problem posing as a study technique. We seek to address the following research questions: 

\begin{itemize}
\item[\textbf{RQ1.}] How do students’ experiences of their interactions with Gen AI for problem-posing in physics change after training on prompt engineering techniques?
\item[\textbf{RQ2.}] What are students' attitudes towards using Gen AI as a tool for practicing problem posing as a method of self-study?
\end{itemize}

\section{Background}

\par Students and educators are beginning to explore the implementation of Gen AI tools, both formally and informally. Gen AI has been employed as a virtual teaching assistant and tutor \cite{hashmi2025chatbot, tang2025TA, ding2023virtualtutor}, as a problem-solving companion \cite{zhu2024humangenerativeaicollaborativeproblem}, and as a lab partner \cite{kilde2025labpartner}, all in a supervised setting. These studies show promising results of the benefit of additional feedback and on-demand guidance from Gen AI. However, when students use Gen AI, they typically do so as a method of self-study outside the classroom, unsupervised.
\par A detailed study by Wang et. al looked into students’ personal uses of Gen AI outside of activities introduced in the classroom \cite{wang2025crutch}. Common uses included explaining concepts from class, summarizing readings, and supporting in-class assignments and homework sets. While many students find Gen AI to be insightful and convenient, it is unclear what the quality of students' interactions with Gen AI are, such as if they are engaging with AI using prompting techniques or other trained skills. It is also unclear if these types of interactions are enhancing their knowledge and problem-solving skills, or simply providing quick answers to complex topics and problems. 
\par Problem posing is a technique that has been popularized in mathematics education \cite{silver1994problemposing, singer2013probposemath,cai2015probposemath,shao2025mathprobposefw}. Rather than solving pre-made problem sets, problem posing asks students to generate meaningful variations to a problem they have previously solved by designing a new problem under specific conditions or modifying a given problem \cite{silver1994problemposing}. One of the greatest hurdles in disciplines requiring complex problem solving is enhancing transfer of learning, i.e. applying knowledge learned to novel scenarios \cite{singley1989transfer}. Developing problem-solving skills requires students to be introduced to methods that prioritize transfer of learning over regurgitating previously solved problems. Problem posing requires students to not only think about how the question can be re-framed, but also the feasibility of the problem in real-world contexts. Problem posing has been to shown to help students overcome the challenges of transfer of learning in mathematics education \cite{akben2020probposmeta, singer2013probposemath, cai2015probposemath}. 



\par The conceptual framework used is based on our prior research \cite{rebello2017transfer} into transfer of learning in problem-solving contexts, which was adapted from Jonassen's framework for problem solving \cite{jonassen2010learning}. We discuss \textit{horizontal transfer}, where the reader takes explicit information from the problem and tries to align it with pre-created knowledge structures, and \textit{vertical transfer}, where a reader actively recognizes features of a situation, activating and applying prior knowledge. In this study, we employ this framework for students to pose problems. Posing new problems via horizontal transfer, i.e., changing a problem's representation or context, might allow students to think more deeply about changes that can be made to problems, while keeping the approach the same. Posing new problems via vertical transfer, i.e., altering the complexity, will require students to apply concepts more deeply, rather than memorizing key examples.
\par One of the challenges of problem-posing instruction is its inherent difficulty as it involves synthesis, which lies at a high level of the revised Bloom's Taxonomy \cite{darwazeh2015revision, krathwohl2002revision}.  Mestre \cite{mestre2002tolprobpose} found that first-year physics students struggled to pose mechanics problems with complete conceptual understanding, even with the scenario provided. Thus, there is a need for scaffolding students' problem posing. 
\par In this study, we prepare students to use Gen AI to provide the scaffolding needed for problem-posing. Gen AI has been shown to scaffold generation of new physics problems for teachers \cite{eladawy2024streamliningprobgeneration} and as a useful source of feedback for a teacher-student problem-posing pedagogy in a middle school mathematics setting \cite{shao2025mathprobposefw}. However, in prior studies, students were not trained with a conceptual framework based on which to pose new problems. In this study, we adapt our theoretically grounded conceptual framework \cite{rebello2017transfer}
to facilitate problem-posing pedagogy, by placing students in the driver's seat, with Gen AI as a guide, to explore problem posing, potentially as a tool for self-study.

\section{Methods}

\subsection{Study and Data Collection}

\par This study was administered as an extra credit assignment in an online first-semester calculus-based physics summer course for prospective engineering majors at a large land grant public university in the U.S. Midwest. The study design consisted of three phases completed over one week. 
\subsubsection{Phase 1 : Pre-Training}
\par In the first phase of the study, students were given limited context on problem posing and the uses of Gen AI. The first prompt asked them to generate a new variation to one of their recitation problems of the week with the help of a Gen AI platform of their choosing. Students were asked to submit their "conversation" with Gen AI and reflect on the interaction with six guided questions. These questions asked students to discuss the similarities and differences between the original and posed problem and the quality of the response produced by Gen AI. They were also asked if they would use the problem posed by Gen AI for studying. 
\subsubsection{Phase 2 : Training}
\par In the second phase of the study, students watched a 15-minute training video, which contained an introduction to Gen AI and how it works. Students were also provided a description of various prompt engineering techniques (from \cite{chen2025prompteng}) and examples of how they are used. The video then introduced our conceptual framework \cite{rebello2017transfer} for problem posing, along with examples of utilizing prompt engineering to pose a physics problem. Students were required to view this video before proceeding to the third phase of the study.
\subsubsection{Phase 3 : Post-Training}
\par In the third phase of the study, students were asked to repeat the problem-posing experiment (phase 1) using what they learned in the training video (phase 2). Students responded to reflection questions that were similar to those in phase 1, with three new questions added. Two new questions regarded the training and techniques used. The final question asked students to reflect on the quality of their interaction with Gen AI before training versus after training and if they would be willing to continue posing problems with Gen AI in the future.

\subsection{Data Analysis}

\par After filtering out incomplete responses, $N = 49$ students completed all three phases of the study, which is approximately 40\% full participation. 
Written open-ended survey responses were analyzed using a descriptive phenomenological approach \cite{morrow2015colaizzi}, in which each response is treated as a participant’s description of how they experienced the phenomenon in question -- use of Gen AI for problem posing. The goal is to identify the meaning and essence of the lived experience as expressed across participants’ responses. All responses were first read holistically, then raters engaged in reflexive memoing to bracket prior assumptions. Significant statements relevant to the phenomenon were identified and clustered into emergent themes. These themes were synthesized into a concise description of shared participant experiences, while preserving variation across individual responses. This is where `codes' for each theme were generated. This approach moves from participant descriptions to thematic structures to form an integrated account of the essence of participants' experiences.
\par To evaluate the validity of our qualitative analysis, three additional raters were recruited to independently categorize each response for the entire data set. Raters were given the context of the study and provided descriptions of the two themes and their codes. 

\begin{table*}[htbp]
\caption{Sample responses from codes for the two themes. The `X' codes are excluded, as responses lacked significant detail.}
\label{table*:1}
\begin{ruledtabular}
\renewcommand{\arraystretch}{2}
\begin{tabular}{@{}l@{\hspace{0.3em}}l@{}}
\textbf{Code} & \textbf{Example Student Response} \\
\hline
1P & \parbox[t]{0.85\textwidth}{\raggedright \textit{In my old prompting techniques, I had a \underline{very basic and solution-focused conversation}. However, with the improved techniques, the experience was \underline{more interactive} compared to the previous one.}} \\
1N & \parbox[t]{0.85\textwidth}{\raggedright \textit{I think my interactions were similar. \underline{I know how to use gen AI}, and I know how to redirect it if needed.}} \\
\hline
2P & \parbox[t]{0.85\textwidth}{\raggedright \textit{I plan to keep using this approach in my future study sessions to \underline{generate targeted practice problems} and get consistently high-quality answers.}} \\
2M & \parbox[t]{0.85\textwidth}{\raggedright \textit{I \underline{wasn't going to use AI} because I thought the questions would be too similar, but \underline{now I can tweak them} in a way that would benefit me.}} \\
2N & \parbox[t]{0.85\textwidth}{\raggedright  \textit{I will absolutely not be using this to study. I have a \underline{strong opposition to the use of generative AI}. Its environmental impact is insane. The amount of harm it does out weighs any sort of benefit. Additionally, I think the \underline{use of AI in academics is a very slippery slope}, and I do not think it should be encouraged.}}\\
\end{tabular}
\end{ruledtabular}
\end{table*}

\vspace{-1em}

\section{Findings \& Discussion}
\label{sec:IV}

\noindent Two themes emerged from our analysis. \textbf{Theme 1} focuses on students’ reports of how of their interactions with Gen AI compare before and after being provided training, specifically their quality. \textbf{Theme 2} focuses on students’ attitudes towards utilizing Gen AI as a tool for problem-posing.

\par Within each of these themes, we coded for types of students' responses. There were three codes for the first theme: \textbf{1P} (positive change), \textbf{1N} (no significant change), and \textbf{1X} (insufficient information). There were four codes for the second theme: \textbf{2P} (positive attitudes), \textbf{2M} (mixed attitudes), \textbf{2N} (negative attitudes), and \textbf{2X} (insufficient information). See Table~\ref{table*:1} for exemplar responses for each code. In some cases, student responses lacked sufficient explanations and did not respond in a way that provided useful insight. This is accounted for in the `insufficient information' codes for each theme. 
\par An average percent agreement and average Cohen’s Kappa value were calculated between the initial rater and each additional rater. For the first theme, the percent agreement was calculated at 80\%, with a Cohen’s Kappa of approximately 0.50. For the second theme, the percent agreement was calculated at 76\%, with a Cohen’s Kappa of approximately 0.53. Thus, the Cohen's Kappa values indicated moderate agreement, falling in the 0.41-0.60 range, between raters for both themes \cite{cohen1960coagree}. In this interpretation, values greater than 0.40 are considered acceptable \cite{cohen1960coagree}.

\subsection{Changes in Interactions Before and After Training}

Based on their perceptions of the interactions, a majority of participants \textbf{(76\%)} perceived a positive change from before to after watching the training video. Of the remaining students, \textbf{13\%} reported no significant change between the pre- and post-training interaction.


\par One of the emergent cross-cutting themes of our analysis is the importance of training in using Gen AI, particularly in prompt engineering techniques. Students who had little to no prior experience with prompt engineering found a significant change in their interactions with Gen AI after the training video. In their reflections on the exercise, students often reported that their interaction with Gen AI prior to training was unfocused, lacked detail, or did not provide enough of a change in their problem to be useful. Post-training, students perceived the interactions as more conversational, relevant, and closely aligned with the prompt. The following student response encapsulates this: 
\vspace{3pt}
\par \textit{"Before using the prompting techniques from the video, my interactions with AI were a bit more basic and sometimes the responses felt less detailed or didn't fully match what I needed. After learning about roleplaying and chain of thought prompting, I noticed the AI's answers became clearer, more thorough, and better explained the steps involved."}
\vspace{3pt}
\par Students who did not perceive a significant change after this training video cited having previous experience interacting with Gen AI platforms or reported using basic prompting techniques, like asking the model to be specific or clear. For example, the following student coded here noted:
\vspace{3pt}
\par \textit{"It helped me put proper names to ideas that I had been using for a long time when it came to AI."}

\subsection{Attitudes on Problem-Posing with Generative AI}

\par Due to the complex nature of the analysis for this theme, responses were sub-categorized by common rationales that students provided in their responses based on their perceptions of the activity.


The results for this theme were generally positive, although more varied than the previous theme. About \textbf{68\%} of students fell into the positive category, expressing a favorable view of the problems posed and the process itself. The following is from a student response rated here:
\vspace{3pt}
\par \textit{"...I will continue using in the future for studying, especially when I need practice problems, conceptual explanations, or tutoring-style feedback in preparation for exams or assignments."}
\vspace{3pt}
\par Similar to the student above, multiple students cited they would utilize this method of study with Gen AI for practice and exam preparation and deepening their understanding of concepts introduced in class or more advanced concepts introduced outside of the classroom. Students also mentioned they would use activities within the course, like quizzes or homework problems, as the basis of posing new problems. An additional set of students found this method helped them to individualize their learning by targeting weak areas, providing feedback, and advancing their problem-solving strategies. 
\par About \textbf{17\%} of students had mixed opinions on the technique and/or the use of Gen AI for self-study. Within the mixed category, students showed interest in using the methods proposed, but cited hesitations. Some students indicate they might use this technique, but it is not their “first choice.” Other students cited their hesitations were due to Gen AI, not fully trusting the generated output or wanting to check the validity of problems posed against other sources. For example, one student coded here noted:
\vspace{3pt}
\par \textit{"...even though this time the problems lacked mistakes and false information, it's not always the case in real life."} 
\vspace{3pt}
\par Some students emphasized wanting to use provided course materials, like practice exams written by an instructor, or other online resources that feel more readily accessible. Another sub-set of students said they found this study method helpful, but might modify the process to fit their needs. For example, one student noted that they achieved the desired results by \textit{"asking for a more accurate and different question"} and \textit{"reasking the same question to see the differences."}
\par About \textbf{9\%} of students held a negative view of the method of problem posing with Gen AI. These students were firm in their oppositional views. Students’ negative perceptions of Gen AI were the main reason cited for their views, including moral and ethical objections to using the tool for coursework and its broader impacts outside of education. Students also tend to distrust Gen AI’s reliability and accuracy in generating realistic problems. Similar to the mixed category, many students tended to prefer existing materials, like those available online or given to them within the course. The following student comments:
\vspace{3pt}
\par \textit{"I still do not enjoy studying with AI because of morality reasons. I do think this is a more beneficial way of using AI than just asking for answers. However, I think that students have gotten this far without generative AI thanks to TAs, textbooks, and the general internet that I feel no need to use AI in my daily study habits."}
\vspace{3pt}
\par It is important to note that students in this category did not express opposition to problem posing, but to using Gen AI's assistance for the practice. Overall, students coded here did not feel the methods or the content created provided benefits greater than the costs of practicing with them.

\subsection{Overarching Trends}

\par Across both themes, a vast majority of students' responses ($\approx72$\%) were positive. Of all the students who perceived a positive change in pre- and post-training interactions (37 of 49), a vast majority of students (27 of 37) also had a positive view of the problem-posing technique with Gen AI. A minority of students (5 of 37) perceived a positive change in interactions, but fell into the mixed rating category with students who showed criticism towards Gen AI and/or its use in problem-posing. Very few students (2 of 49) were positive about their change in interactions after training but negative about the use Gen AI for problem posing, and fewer still (1 of 49) were negative about both aspects. Students who found a positive change in their interaction with Gen AI but had a negative perception of problem-posing with Gen AI, still strengthen the claim that training on prompt engineering is crucial for students who plan to engage with Gen AI. The student who found no significant change in their interactions and did not favor problem-posing with Gen AI described their negative perceptions of the implications of using Gen AI.

\section{Conclusions \& Implications}

\par This work adds to the body of literature investigating the use of Gen AI in physics education. Simultaneously, it addresses the use of problem posing -- an established yet understudied technique in STEM education -- as a self-study technique and recruiting Gen AI to facilitate its application. 
\par \textbf{RQ 1.} inquired about changes in students' experiences of their interactions with Gen AI for problem posing after training. We found that the overwhelming majority of students experienced a positive change in their interaction with Gen AI before and after receiving training on prompt engineering techniques. Those who did not experience a positive change were generally familiar with some of these techniques. This work supports key findings from Generative AI-related studies \cite{chen2025prompteng,wang2025crutch}, emphasizing the importance of prompt engineering training for students using Generative AI platforms in these ways. 
\par \textbf{RQ 2.} inquired about students’ attitudes towards using Gen AI as a tool for practicing problem-posing as a method of self-study. We found that students' perceptions of problem posing with Gen AI remain generally positive, with some students reporting that they are willing to adapt this method of study to fit their individual needs. Students who held reservations to problem posing preferred more readily available learning materials, such as those provided in the course. Some students held reservations regarding the use of Gen AI in any context. 

\section{Limitations \& Future Work}
\par The online modality and size of the course made it difficult to recruit many participants. Students actively chose to participate in this assignment for extra credit, which might have skewed views to be more agreeable. Conversely, students opposed to the use of Gen AI may have chosen to not participate, meaning some valuable results might have been lost. Additionally, qualitative analysis relied on each rater to determine the perceptions of students based on their answers to the reflection questions, instead of interview-style accounts. This may have contributed to a lower Cohen's Kappa (< 0.6).
\par This study opens the door for further investigation of incorporating Gen AI into physics education, with the openness of the majority of students to incorporating Gen AI into structured self-study techniques. The results of this study additionally support the introduction of Gen AI training into the curricula. This lays the foundation for an examination of training and how to employ methods widely, keeping in mind the subset of students who hold moral objections to the use of AI. Future research with this data set will examine the ‘conversations’ students had, both before and after their training, to assess the prompt engineering methods used and to evaluate the problems posed by them. This study will provide a more direct look at students’ interactions with Gen AI with our problem-posing technique and will address the limitations associated with analyzing student perceptions. 

\section{Acknowledgments}
This work is supported in part by a U.S. National Science Foundation grant 2111138. Opinions expressed are of the authors and not of the Foundation.

\newpage

\bibliographystyle{apsrev4-2}
\bibliography{references}

\end{document}